\documentclass[11pt]{article}

\makeatletter%
\def\nottoobig#1{{\hbox{$\left#1\vcenter to1.111\ht\strutbox{}\right.\n@space$}}}
\makeatother%

\makeatother%

\makeatletter%

\newcount\hour  \newcount\minutes  \hour=\time  \divide\hour by 60
\minutes=\hour  \multiply\minutes by -60  \advance\minutes by \time
\def\mmmddyyyy{\ifcase\month\or Jan\or Feb\or Mar\or Apr\or May\or Jun\or Jul\or
  Aug\or Sep\or Oct\or Nov\or Dec\fi \space\number\day, \number\year}
\def\hhmm{\ifnum\hour<10 0\fi\number\hour :%
  \ifnum\minutes<10 0\fi\number\minutes}
\def\Draft{{\it Draft of \mmmddyyyy}}

\def\ps@jtsheadings{%
\def\@oddhead{\it\rightmark\hfil\rm\thepage}%
\def\@oddfoot{\hfil\Draft}%
\if@twoside%
\def\@evenhead{\rm\thepage\hfil\it\leftmark}%
\def\@evenfoot{\Draft\hfil}%
\else
\let\@evenhead\@oddhead%
\let\@evenfoot\@oddfoot%
\fi%
}
\def\ps@jtsplain{%
\def\@oddhead{\hfil\Draft}%
\def\@oddfoot{\hfil\rm\thepage\hfil}%
\let\@evenfoot\@oddfoot%
\if@twoside \def\@evenhead{\Draft\hfil} \else \let\@evenhead\@oddhead \fi
}

\def\chaptermark#1{\markboth{\thechapter.\ #1}{\thechapter.\ #1}}%
\def\sectionmark#1{\markright{\thesection.\ #1}}

\def\section{\@startsection {section}{1}{\z@}
    {3.5ex plus1ex minus.2ex}{2.3ex plus.2ex}{\Large\bf}}
\def\subsection{\@startsection{subsection}{2}{\z@}
    {3.25ex plus1ex minus.2ex}{1.5ex plus.2ex}{\large\bf}}
\def\subsubsection{\@startsection{subsubsection}{3}{\z@}
    {3.25ex plus1ex minus.2ex}{1.5ex plus.2ex}{\normalsize\bf}}
\def\paragraph{\@startsection{paragraph}{4}{\z@}
    {3.25ex plus1ex minus.2ex}{1em}{\normalsize\bf}}
\def\subparagraph{\@startsection{subparagraph}{4}{\parindent}
    {3.25ex plus1ex minus.2ex}{1em}{\normalsize\bf}}

\makeatother%

\makeatletter \@beginparpenalty=10000 \makeatother

\def\underl#1 {\leavevmode\let\first=\relax\underli #1 }
\def\underli#1 {\ifx&#1\let\next=\relax\unskip
                \else\let\next=\underli\first\ulinebox{#1}\fi\let\first=\undersp\next}
\def\undersp{\penalty50\ulinebox{\space}\penalty50}
\def\ulinebox#1{\vtop{\hbox{\strut#1}\hrule}}%
\def\unice#1 {\underl #1 & }
\def\desclabel#1{\bf #1\hfil}
\def\desc{\list{}{%
\labelwidth=\leftmargin
\advance \labelwidth by -\labelsep
\let \makelabel=\desclabel}}

\makeatletter %

\newcommand{\implies}{\:\Rightarrow\:}

\def\union{\,\bigcup\limits\,}

\newlength{\filength}
\newsavebox{\gcbox}
\sbox{\gcbox}{\framebox[\filength]{\rule{0ex}{2ex}}}

\newlength{\leftjustindent}
\newlength{\@leftjustindent}
\def\leftjust{\let\\\@leftjustcr\let\end\@endleftjust
  \addtolength{\@leftjustindent}{\leftjustindent}
  \vcenter\bgroup
  \halign\bgroup
    \hbox to\displaywidth{
      \rule{\@leftjustindent}{0ex}$\displaystyle##$\hfill
      }\crcr
}
\def\endleftjust{\crcr\egroup\egroup\endgroup}
\def\@endleftjust#1{\crcr\egroup\egroup\@checkend{#1}\endgroup}
\def\@leftjustcr{\crcr}

\newtheorem{theorem}{Theorem}[section]
\newtheorem{corollary}[theorem]{Corollary}
\newtheorem{definition}[theorem]{Definition}

\newcommand{\qedblob}{\mbox{\rule[-1.5pt]{5pt}{10.5pt}}}
\def\literalqed{{\ \nolinebreak\hfill\mbox{\qedblob\quad}}}

\def\qed{\literalqed}

\newtheorem{lemma}[theorem]{Lemma}

\newcommand{\singlespacing}{\let\CS=
\@currsize\renewcommand{\baselinestretch}{1}\tiny\CS}
\newcommand{\singlespacingplus}{\let\CS=
\@currsize\renewcommand{\baselinestretch}{1.25}\tiny\CS}
\newcommand{\doublespacing}{\let\CS=
\@currsize\renewcommand{\baselinestretch}{1.75}\tiny\CS}
\newcommand{\draftspacing}{\let\CS=
\@currsize\renewcommand{\baselinestretch}{2.0}\tiny\CS}

\makeatother%

\mathcode`\0="0030      %
\mathcode`\1="0031
\mathcode`\2="0032
\mathcode`\3="0033
\mathcode`\4="0034
\mathcode`\5="0035
\mathcode`\6="0036
\mathcode`\7="0037
\mathcode`\8="0038
\mathcode`\9="0039
\singlespacingplus
\makeatletter
\clubpenalty=\@highpenalty
\widowpenalty=\@highpenalty
\makeatother

\makeatletter
\newcommand{\niceonespacing}{\let\CS=\@currsize\renewcommand{\baselinestretch}{1.1}\tiny\CS}\newcommand{\nicetwospacing}{\let\CS=\@currsize\renewcommand{\baselinestretch}{1.2}\tiny\CS}
\newcommand{\nicethreespacing}{\let\CS=\@currsize\renewcommand{\baselinestretch}{1.3}\tiny\CS}
\newcommand{\singlespacingplusplus}{\let\CS=\@currsize\renewcommand{\baselinestretch}{1.35}\tiny\CS}
\newcommand{\nicefourspacing}{\let\CS=\@currsize\renewcommand{\baselinestretch}{1.4}\tiny\CS}
\newcommand{\nicefivespacing}{\let\CS=\@currsize\renewcommand{\baselinestretch}{1.5}\tiny\CS}
\newcommand{\nicesixpacing}{\let\CS=\@currsize\renewcommand{\baselinestretch}{1.6}\tiny\CS}
\makeatother

\makeatletter
\def\@cite#1#2{[#1\if@tempswa , #2\fi]}
\makeatother

\makeatletter
\def\@citex[#1]#2{\if@filesw\immediate\write\@auxout{\string\citation{#2}}\fi
  \def\@citea{}\@cite{\@for\@citeb:=#2\do
    {\@citea\def\@citea{,\linebreak[0]}\@ifundefined
       {b@\@citeb}{{\bf ?}\@warning
       {Citation `\@citeb' on page \thepage \space undefined}}%
\hbox{\csname b@\@citeb\endcsname}}}{#1}}
\makeatother

\makeatletter
\def\ps@thesis{\def\@oddhead{\hfil\rm\thepage\hfil}\def\@oddfoot{}\def\@evenhead{\hfil\rm\thepage\hfil}\def\@evenfoot{}\def\chaptermark##1{}\def\sectionmark##1{}}
\makeatother

\makeatletter
\def\foobarpt{\textfont\z@\tenrm 
  \scriptfont\z@\ninrm \scriptscriptfont\z@\sevrm
\textfont\@ne\tenmi \scriptfont\@ne\ninmi \scriptscriptfont\@ne\sevmi
\textfont\tw@\tensy \scriptfont\tw@\ninsy \scriptscriptfont\tw@\sevsy
\textfont\thr@@\tenex \scriptfont\thr@@\tenex \scriptscriptfont\thr@@\tenex
\def\unboldmath{\everymath{}\everydisplay{}\@nomath\unboldmath
          \textfont\@ne\tenmi 
          \textfont\tw@\tensy \textfont\lyfam\tenly
          \@boldfalse}\@boldfalse
\def\boldmath{\@ifundefined{tenmib}{\global\font\tenmib\@mbi\@magscale1\global
        \font\tensyb\@mbsy \@magscale1\global\font
         \tenlyb\@lasyb\@magscale1\relax\@addfontinfo\@xiipt
              {\def\boldmath{\everymath
                {\mit}\everydisplay{\mit}\@prtct\@nomathbold
                \textfont\@ne\tenmib \textfont\tw@\tensyb 
                \textfont\lyfam\tenlyb\@prtct\@boldtrue}}}{}\@xiipt\boldmath}%
\def\prm{\fam\z@\tenrm}%
\def\pit{\fam\itfam\tenit}\textfont\itfam\tenit \scriptfont\itfam\ninit
   \scriptscriptfont\itfam\sevit
\def\psl{\fam\slfam\tensl}\textfont\slfam\tensl 
     \scriptfont\slfam\tensl \scriptscriptfont\slfam\tensl
\def\pbf{\fam\bffam\tenbf}\textfont\bffam\tenbf 
   \scriptfont\bffam\ninbf \scriptscriptfont\bffam\ninbf 
\def\ptt{\fam\ttfam\tentt}\textfont\ttfam\tentt
   \scriptfont\ttfam\nintt \scriptscriptfont\ttfam\nintt 
\def\psf{\fam\sffam\tensf}\textfont\sffam\tensf
    \scriptfont\sffam\tensf \scriptscriptfont\sffam\tensf
\def\psc{\@getfont\psc\scfam\@xiipt{\@mcsc\@magscale1}}%
\def\ly{\fam\lyfam\tenly}\textfont\lyfam\tenly 
   \scriptfont\lyfam\ninly \scriptscriptfont\lyfam\sevly
 \@setstrut \rm}

\makeatother

\newcommand{\p}{{\rm P}}

\newcommand{\littlepitalic}{{\it p}}

\newcommand{\np}{{\rm NP}}

\newcommand{\bh}{{\rm BH}}

\newcommand{\sigmaj}{{{\rm \Sigma}_j^p}}

\newcommand{\sigmazero}{{{\rm \Sigma}_{0}^p}}
\newcommand{\sigmaone}{{{\rm \Sigma}_{1}^p}}
\newcommand{\sigmakminusone}{{{\rm \Sigma}_{k-1}^p}}
\newcommand{\sigmakminustwo}{{{\rm \Sigma}_{k-2}^p}}
\newcommand{\sigmak}{{{\rm \Sigma}_{k}^p}}

\newcommand{\sigmakplusone}{{\rm \Sigma}_{k+1}^p}
\newcommand{\sigmakmone}{{\rm \Sigma}_{k-1}^p}
\newcommand{\sigmakmtwo}{{\rm \Sigma}_{k-2}^p}
\newcommand{\sigmai}{{\rm \Sigma}_i^p}

\newcommand{\pik}{{\rm \Pi}_k^p}
\newcommand{\pikmone}{{\rm \Pi}_{k-1}^p}
\newcommand{\deltai}{{\rm \Delta}_i^p}

\newcommand{\deltak}{{\rm \Delta}_k^p}
\newcommand{\deltakmone}{{\rm \Delta}_{k-1}^p}

\newcommand{\psigkmqueries}{ {\p^{\sigmak[m]}}}
\newcommand{\psigkmplusone}{ {\p^{\sigmak[m+1]}}}
\newcommand{\wh}[1]{\widehat{#1}}

\newcommand{\lsigkprime}{ {L'_{\sigmak}}}
\newcommand{\lsigi}{ {L_{\sigmai}}}

\newcommand{\lsigkmone}{ {L_{\sigmakmone}}}

\newcommand{\lsigkmtwoprimeprime}{ {L''_{\sigmakmtwo}}}

\newcommand{\ldelkmone}{ {L_{\deltakmone}}}

\newcommand{\psigione}{ {\p^{\sigmai[1]}}}

\newcommand{\psigjone}{ {\p^{\sigmaj[1]}}}

\newcommand{\diffmsigk}{{\rm DIFF}_m(\sigmak)}
\newcommand{\diffssigi}{{\rm DIFF}_s(\sigmai)}

\newcommand{\diffmplusonesigk}{{\rm DIFF}_{m+1}(\sigmak)}
\newcommand{\diff}{{\rm DIFF}}
\newcommand{\codiffmsigk}{{\rm co}{\diffmsigk}}
\newcommand{\codiffmplusonesigk}{{\rm co}{\diffmplusonesigk}}
\newcommand{\ldiffmsigk}{{L_{\diffmsigk}}}

\newcommand{\lhatdiffmsigk}{{\wh{L}_{\diffmsigk}}}
\newcommand{\deltatilde}{\tilde{\Delta}}
\newcommand{\bolddelta}{{\bf \Delta}}

\newcommand{\psigmakone}{  {\p^{ {\rm \Sigma_{\it k}^{\it p}}[1]}}}
\newcommand{\psigmaktwo}{  {\p^{ {\rm \Sigma_{\it k}^{\it p}}[2]}}}

\newcommand{\conp}{{\rm coNP}}

\newcommand{\sigmatwo}{{\Sigma_2^p}}
\newcommand{\pitwo}{{\Pi_2^p}}

\newcommand{\ph}{{\rm PH}}

\newcommand{\proof}{\noindent {\bf Proof:}\quad}

\singlespacingplus

\def\pair#1{{{\langle\!\!~#1~\!\!\rangle}}}

\newcommand{\manyone}{\mbox{$\,\leq_{\rm m}^{{\littlepitalic}}$\,}}

\newcommand{\sigmastar}{\mbox{$\Sigma^\ast$}}
\newcommand{\pisnp}{\mbox{$\p=\np$}}

\newcommand{\calc}{{\cal C}}

\newcommand{\condition}{\,\nottoobig{|}\:}

\singlespacingplus

\title{%
Translating Equality Downwards}

\author{
Edith Hemaspaandra\thanks{%
Email: {\tt edith@bamboo.lemoyne.edu}.
Supported in part 
by grant
NSF-INT-9513368/DAAD-315-PRO-fo-ab.  Work done in part while 
visiting 
Friedrich-Schiller-Universit\"at Jena.}
\\Department of Mathematics\\
Le Moyne College\\
Syracuse, NY 13214, USA
\and
Lane A. Hemaspaandra\thanks{%
Email: {\tt lane@cs.rochester.edu}.
Supported in part 
by grants NSF-CCR-9322513 and 
NSF-INT-9513368/DAAD-315-PRO-fo-ab.  Work done in part while 
visiting 
Friedrich-Schiller-Universit\"at Jena.}
\\Department of Computer Science\\University of Rochester\\
            Rochester, NY 14627, USA
\and 
Harald Hempel\thanks{%
Email: {\tt hempel@mipool.uni-jena.de}.
Supported in part 
by grant
NSF-INT-9513368/DAAD-315-PRO-fo-ab.  Work done in part 
while visiting Le~Moyne College.}
\\Inst.~f\"ur Informatik\\
Friedrich-Schiller-Universit\"at Jena\\
07743 Jena, Germany}

\date{January 31, 1998}
\makeatletter
\def\@listI{\leftmargin\leftmargini \parsep 4.5pt plus 1pt minus 1pt\topsep
6pt plus 2pt minus 2pt \itemsep  2pt plus 2pt minus 1pt}

\let\@listi\@listI
\@listi
\makeatother

\begin{document}

\typeout{PLEASE NEVER DELETE COMMENTED OUT STUFF!! It is part of the paper's memory.  Cheers, Lane}
\typeout{PLEASE NEVER DELETE COMMENTED OUT STUFF!! It is part of the paper's memory.  Cheers, Lane}
\typeout{PLEASE NEVER DELETE COMMENTED OUT STUFF!! It is part of the paper's memory.  Cheers, Lane}

\typeout{WARNING:  BADNESS used to suppress reporting!  Beware!!}
\hbadness=3000%
\vbadness=10000 %

\bibliographystyle{alpha}

\pagestyle{empty}
\setcounter{page}{1}

\pagestyle{empty}
\setcounter{footnote}{0}
{\singlespacing\maketitle}

\begin{abstract}
{\singlespacing

Downward translation of equality refers to cases 
where a collapse of some pair of 
complexity classes would induce a collapse
of some other pair of complexity classes that (a priori) one expects
are smaller.  Recently,
the first downward translation of equality was obtained that 
applied to the polynomial hierarchy---in particular, to 
bounded access to its levels~\cite{hem-hem-hem:cOutByJour:downward-translation}.  
In this paper, we provide a much broader downward translation that 
extends not only that downward translation but also 
that translation's elegant enhancement by Buhrman
and Fortnow~\cite{buh-for:t:two-queries}.  Our work also sheds 
light on previous research
on the structure
of refined polynomial 
hierarchies~\cite{sel:j:fine-hierarchies,sel:c:refined-ph},
and strengthens the connection between the collapse of 
bounded query hierarchies and the collapse of the 
polynomial hierarchy.

} %
\end{abstract}

\singlespacing
\setcounter{page}{1}
\pagestyle{plain}
\sloppy

\section{Introduction}

Does the collapse of low-complexity classes imply the collapse of
higher-complexity classes?  Does the collapse of high-complexity
classes imply the collapse of lower-complexity classes?  These
questions---known respectively as downward and 
upward translation of
equality---have long been central topics in computational complexity
theory.  For example, in the seminal paper on the polynomial
hierarchy, Meyer and Stockmeyer~\cite{mey-sto:c:reg-exp-needs-exp-space}
proved that the polynomial hierarchy displays upward translation of
equality (e.g., $\pisnp \implies \p = \ph$).  

The issue of whether the polynomial hierarchy---its levels and/or
bounded access to its levels---ever displays {\em downward\/}
translation of equality has proved more difficult.  The first such
result was recently obtained by Hemaspaandra, Hemaspaandra, and
Hempel~\cite{hem-hem-hem:cOutByJour:downward-translation}, 
who proved that if for some high level of the polynomial
hierarchy one query equals two queries, then the hierarchy collapses
down not just to one query to that level, but rather to that level
itself.  That is, they proved the following result (note: 
the levels  of the polynomial 
hierarchy~\cite{mey-sto:c:reg-exp-needs-exp-space,sto:j:poly}
are denoted in the standard way, namely,
$\sigmazero = \p$, $\sigmaone = \np$, 
$\sigmak = \np^{\sigmakminusone}$ for each $k>1$, and 
$\pik = \{ L \condition \overline{L} \in \sigmak\}$ for 
each $k \geq 0$).

\begin{theorem}\label{t:hhh-nonBHdownward}
(\cite{hem-hem-hem:cOutByJour:downward-translation}) 
\quad
For each $k>2$: If $\psigmakone = \psigmaktwo$, then
$\sigmak  = \pik = \ph$.  
\end{theorem}

This theorem has two clear directions in which one might hope to
strengthen it.  First, one might ask 
not just about one-versus-two queries but
rather about $j$-versus-$j+1$ queries.  Second, one might ask if the
$k>2$ can be improved to $k>1$.  Both of these have been achieved.
The first strengthening was achieved in a more technical section of
the same paper by Hemaspaandra, Hemaspaandra, and
Hempel~\cite{hem-hem-hem:cOutByJour:downward-translation}. 
They showed that
Theorem~\ref{t:hhh-nonBHdownward} was just the $j=1$ special case of a
more general downward translation result they established, for $k>2$,
between bounded access to $\sigmak$ and the boolean hierarchy over
$\sigmak$.  The second type of strengthening was achieved by Buhrman
and Fortnow~\cite{buh-for:t:two-queries}, who in a very
elegant paper showed that
Theorem~\ref{t:hhh-nonBHdownward} holds even for $k=2$, but who also
showed that no relativizable technique can establish
Theorem~\ref{t:hhh-nonBHdownward} for $k=1$.

Neither of the results or proofs just mentioned is broad enough to
achieve both strengthenings simultaneously.  In this paper we present
new results strong enough to achieve this---and more.  In
particular, we unify and extend all the above results, and also unify
with these results 
and extend 
the most computer-science-relevant portions of the work of 
Selivanov
\mbox{(\cite[Section~8]{sel:j:fine-hierarchies},\cite{sel:c:refined-ph})}
on whether refined
polynomial hierarchy classes are closed under complementation.  

To explain exactly what we do and how it extends previous results, we
now state the abovementioned results in the more general forms in
which they were actually established, though in some cases with 
different notations or statements (see, e.g., the 
interesting recent paper of Wagner~\cite{wag:t:parallel-difference}
regarding the relationship between ``delta notation''
and truth-table classes).  
Before stating the results, we must very briefly
remind the reader of three definitions/notations, namely 
of the $\Delta$ levels of the polynomial hierarchy, 
of symmetric difference, and of 
boolean hierarchies.

\begin{definition}
\begin{enumerate}

\item (see \cite{mey-sto:c:reg-exp-needs-exp-space}) \quad  As is standard,
for each $k \geq 1$, $\deltak$ denotes
$\p^{\sigmakminusone}$.  

\item
For any classes ${\cal C}$ and ${\cal D}$, 
\[{\cal C} \bolddelta {\cal D} = \{L \condition (\exists C \in {\cal C})
(\exists D \in {\cal D}) [ L = C \Delta D]\},\]
where $C \Delta D = (C - D) \cup (D - C)$.

\item (\cite{cai-gun-har-hem-sew-wag-wec:j:bh1,cai-gun-har-hem-sew-wag-wec:j:bh2}, see also~\cite{hau:b:sets,koe-sch-wag:j:diff}) \quad
Let $\cal C$ be any complexity class.  We now define the levels of 
the boolean hierarchy.
\begin{enumerate}
\item $ {\rm DIFF}_1(\calc) = \calc$.
\item For any $k \geq 1$, ${\rm DIFF}_{k+1}(\calc) =
\{ L \condition  (\exists L_1 \in \calc)(\exists L_2 \in 
{\rm DIFF}_k(\calc))[ L = L_1 - L_2]\}$.
\item For any $k \geq 1$, $ {\rm coDIFF}_k(\calc) = 
\{ L \condition \overline{L} \in 
{\rm DIFF}_k(\calc)\}$.
\item $\bh(\calc)$, the boolean hierarchy over $\calc$, is 
$\union_{k \geq 1\,} {\rm DIFF}_k$.
\end{enumerate}
\end{enumerate}
\end{definition}
The relationship between the levels of the boolean 
hierarchy over $\sigmak$ and bounded access 
to $\sigmak$ is as follows.  
For each $k \geq 0$ and each $m \geq 0$,
$\p^{\sigmak[m]} 
{
{\subseteq \, \diffmplusonesigk  \, \subseteq}
\atop
{\subseteq  \, \codiffmplusonesigk  \, \subseteq}
}
\p^{\sigmak[m+1]}$.

Now we can state what the earlier papers achieved (and, in doing so,
those papers obtained as corollaries the results mentioned above).

\begin{theorem}\label{t:three-things}
\begin{enumerate}
\item \label{p:hhh-BHdown}
(\cite{hem-hem-hem:cOutByJour:downward-translation}) \quad
Let $m > 0$, $0 \leq i < j < k$, and $ i <
k-2$. If $\psigione \bolddelta \diffmsigk = \psigjone \bolddelta \diffmsigk$,
then 
$\diffmsigk = \codiffmsigk$.
\item (\cite{buh-for:t:two-queries}) \quad If $\p \bolddelta \sigmatwo =  \np
\bolddelta \sigmatwo$, then
$\sigmatwo  = \pitwo = \ph.$
\item (\cite{sel:j:fine-hierarchies,sel:c:refined-ph}) \quad 
If $\sigmai \bolddelta \sigmak$ is closed
under complementation, then the polynomial hierarchy 
collapses.\footnote{Selivanov
\protect\cite{sel:j:fine-hierarchies,sel:c:refined-ph}
establishes only that the hierarchy collapses to a higher 
level, namely a level that contains $\sigmakplusone$; thus this result is an 
upward translation of equality rather than a downward
translation of equality.}
\end{enumerate}
\end{theorem}

In this paper, we unify all three of the above results---and 
achieve the strengthened corollary alluded to above
(and stated later as Corollary~\ref{c:special-case}) 
regarding the relative power
of $j$ and $j+1$ queries to 
$\sigmak$---by
proving the following two results, each of which is a 
downward translation of equality.

\begin{enumerate}
\item Let $m > 0$ and $0 < i < k$. 
If $\deltai \bolddelta \diffmsigk = \sigmai
\bolddelta \diffmsigk$, then
$\diffmsigk = \codiffmsigk$.
\item
Let $m > 0$ and $0 < i < k - 1$.
If $\sigmai \bolddelta \diffmsigk$ is closed under complementation,
then $\diffmsigk = \codiffmsigk$.
\end{enumerate}

Informally put, the technical innovation of our proof is as follows.  In
the previous work extending 
Theorem~\ref{t:hhh-nonBHdownward}
to the boolean hierarchy (part~\ref{p:hhh-BHdown} of
Theorem~\ref{t:three-things}), the ``coordination'' difficulties
presented by the fact that boolean hierarchy sets are in 
effect handled via collections of machines were resolved via 
using certain lexicographically extreme objects as clear signposts
to signal machines with.  In the current stronger context that 
approach fails.  Instead, we integrate into the 
structure of easy-hard-technique proofs
(especially those 
of~\cite{hem-hem-hem:cOutByJour:downward-translation,buh-for:t:two-queries})
the so-called ``telescoping'' normal form possessed by the 
boolean hierarchy over $\sigmak$ (for 
each $k$)~\cite{cai-gun-har-hem-sew-wag-wec:j:bh1,hau:b:sets,wec:c:bh:ormaybe:wech:only:is:right}, 
which in concept dates back to Hausdorff's work on
algebras of sets.  This normal form
guarantees that if $L \in \diffmsigk$, then 
there are sets 
$L_1, L_2, \cdots, L_m \in \Sigma^p_k$ such that
$\ldiffmsigk = L_1 - (L_2 - (L_3 - \cdots  (L_{m-1} - L_m) \cdots))$
and $L_1 \supseteq L_{2} \supseteq \cdots \supseteq L_{m-1}
\supseteq L_m$.  (Picture, if you will, an archery target with 
concentric rings of membership and nonmembership.  That is exactly
the effect created by this normal form.)  

As noted at the end of Section~\ref{s:conclusions},
the stronger downward translations we obtain yield a 
strengthened collapse of the polynomial hierarchy under the
assumption of a collapse in the bounded query hierarchy
over $\np^{\rm NP}$.

We conclude this section with 
some additional literature pointers.
We mention that the proofs of Theorem~\ref{t:hhh-nonBHdownward}
and all that grew out of it---including this paper---are indebted to,
and use extensions of, the ``easy-hard'' technique that was 
invented by 
Kadin (\cite{kad:joutdatedbychangkadin:bh}, 
as further developed 
in~\cite{wag:t:n-o-q-87version,wag:t:n-o-q-89version,bei-cha-ogi:j:difference-hierarchies,cha-kad:j:closer})
to study {\em upward\/} translations of equality resulting 
from the collapse of the boolean hierarchy.
We also mention that there is a body of 
literature showing that equality of 
exponential-time classes translates downwards
in a 
limited sense:  Relationships are obtained with 
whether {\em sparse\/} sets collapse within lower 
time classes (the classic paper in this area
is that of Hartmanis, Immerman, 
and Sewelson~\cite{har-imm-sew:j:sparse},
see also~\cite{rao-rot-wat:j:upward};
limitations of such results
are presented in~\cite{all:j:lim,all-wil:j:downward,hem-jha:j:defying}).
Other than being a restricted type of downward translation
of equality, that body of work has no close connection with the present
paper due to that body of 
work's applicability only to sparse sets.

\section{Main Result: A New Downward Translation of Equality}

We first need a definition and a useful lemma.

\begin{definition}
For any sets ${C}$ and ${D}$:
\[C \deltatilde D =
\{\pair{x, y} \condition x \in C \Leftrightarrow y \not
\in D\}.\]
\end{definition}

\begin{lemma}\label{l:tildecomplete}
$C$ is $\manyone$-complete for ${\cal C}$ and
$D$ is $\manyone$-complete for ${\cal D}$, then
$C \deltatilde D$ is $\manyone$-hard for ${\cal C} \bolddelta {\cal D}$.
\end{lemma}
\proof
Let $L \in {\cal C} \bolddelta {\cal D}$. We need to show that $L \manyone 
C \deltatilde D$.  Let $\wh{C} \in {\cal C}$ and $\wh{D} \in {\cal D}$ be such
that $L = \wh{C} \Delta \wh{D}$. Let $\wh{C} \manyone C$ by $f_C$,
and $\wh{D} \manyone D$ by $f_D$.  
Then $x \in L$ iff $x \in \wh{C} \Delta \wh{D}$,
$x \in \wh{C} \Delta \wh{D}$
iff $(x \in \wh{C} \Leftrightarrow x
\not \in \wh{D})$,
$(x \in \wh{C} \Leftrightarrow x
\not \in \wh{D})$
iff 
$(f_C(x) \in C \Leftrightarrow f_D(x)
\not \in D)$,
and
$(f_C(x) \in C \Leftrightarrow f_D(x)
\not \in D)$
iff $\pair{f_C(x), f_D(x)} \in C \deltatilde D$.~\qed

We now state our main result.

\begin{theorem}\label{t:globalmain}
Let $m > 0$ and $0 < i < k$. If $\deltai \bolddelta \diffmsigk = \sigmai
\bolddelta \diffmsigk$, then
$\diffmsigk = \codiffmsigk$.
\end{theorem}

This result almost follows 
from the forthcoming Theorem~\ref{t:rest}---or, to be more 
accurate, almost all of its cases are easy corollaries of 
Theorem~\ref{t:rest}.
However, the remaining cases---which are the 
most challenging ones---also need 
to be established, and Theorem~\ref{t:keysubpart} does 
exactly that.

\begin{theorem}\label{t:keysubpart}
Let $m >  0$ and  $k>1$.
If $\deltakmone \bolddelta \diffmsigk = \sigmakmone \bolddelta \diffmsigk$,
then $\diffmsigk = \codiffmsigk$.
\end{theorem}

\begin{definition}\label{d:languages}
For each $ k>1$, choose any fixed problem
that is $\manyone$-complete for $\sigmak$
and call it 
$\lsigkprime$.  Now, having fixed such sets, 
for each 
$k>1$ choose one fixed set  $\lsigkmtwoprimeprime$ that 
is in $\Sigma_{k-2}^p$ and one fixed set $\lsigkmone$
that is $\manyone$-complete for $\Sigma_{k-1}^p$
and 
that satisfy\footnote{By 
Stockmeyer's~\cite{sto:j:poly} 
standard quantifier characterization
of the polynomial hierarchy's levels, there do exist 
sets
satisfying this definition.}
$$\lsigkprime = \{{x} \condition
(\exists y \in \Sigma^{|{x}|}) (\forall z \in \Sigma^{|{x}|})
[\pair{{x},y,z} \in  \lsigkmtwoprimeprime]\}$$
and
$$\lsigkmone = \{\pair{{x},y,z} \condition |{x}| = |y|
\wedge (\exists z') [(|{x}| = |y| = |zz'|) \wedge
\pair{{x}, y, zz'} \not\in \lsigkmtwoprimeprime]\}.$$
\end{definition}

\smallskip

\noindent
{\bf Proof of Theorem~\ref{t:keysubpart}} \quad
Let $\lsigkmone \in \sigmakmone$ be as defined in
Definition~\ref{d:languages}, and let $\ldelkmone$ and
$\ldiffmsigk$ be any fixed $\manyone$-complete sets for
$\deltakmone$ and $\diffmsigk$, respectively;  such languages exist, e.g., via
the standard canonical complete set  constructions using enumerations of clocked
machines. {}From Lemma~\ref{l:tildecomplete} it follows that 
$\ldelkmone \deltatilde \ldiffmsigk$ is \manyone-hard for
$\deltakmone \bolddelta \diffmsigk$.
(Though this is not needed for this proof, we note in
passing that it also 
can be easily seen to be in
$\deltakmone \bolddelta \diffmsigk$, and so 
it is in fact $\manyone$-complete for
$\deltakmone \bolddelta \diffmsigk$.)
Since 
$\lsigkmone \deltatilde \ldiffmsigk \in \sigmakmone \bolddelta \diffmsigk$
and by assumption
$\deltakmone \bolddelta \diffmsigk = \sigmakmone \bolddelta \diffmsigk$,
there exists a polynomial-time many-one reduction $h$ from
$\lsigkmone \deltatilde \ldiffmsigk$ to
$\ldelkmone \deltatilde \ldiffmsigk$ (in light of 
the latter's $\manyone$-hardness).
So, for all $x_1, x_2 \in \sigmastar$:
if $h(\pair{x_1, x_2}) = \pair{y_1, y_2}$, 
then
$(x_1 \in \lsigkmone \Leftrightarrow x_2 \not\in \ldiffmsigk)$
if and only if 
$(y_1 \in \ldelkmone \Leftrightarrow y_2 \not\in \ldiffmsigk)$.
Equivalently, 
for all $x_1, x_2 \in \sigmastar$:
\begin{quotation}
\noindent
if $h(\pair{x_1, x_2}) = \pair{y_1, y_2}$, \\ 
then
$$(x_1 \in \lsigkmone \Leftrightarrow x_2 \in \ldiffmsigk)
\mbox{ if and only if }
(y_1 \in \ldelkmone \Leftrightarrow y_2 \in \ldiffmsigk).\ \ 
\hfil \hfill (**) $$
\end{quotation}

We can use $h$ to recognize some of $\overline{\ldiffmsigk}$ by a $\diffmsigk$
algorithm. 
In particular, 
we say that a string $x$ is {\em
easy for length $n$\/} if there exists a string $x_1$ such that
$|x_1| \leq n$ and $(x_1 \in \lsigkmone
\Leftrightarrow y_1 \not \in \ldelkmone)$ where
$h(\pair{x_1, x}) = \pair{y_1, y_2}$.

Let $p$ be a fixed polynomial, which will be exactly 
specified later in the proof.  We have the following algorithm to test whether
$x \in \overline{\ldiffmsigk}$ in the 
case that (our input) $x$ is an easy string
for $p(|x|)$.
Guess $x_1$ with $|x_1| \leq p(|x|)$, let
$h(\pair{x_1, x}) = \pair{y_1, y_2}$,
and accept if and only if
$(x_1 \in \lsigkmone \Leftrightarrow y_1 \not \in \ldelkmone)$ 
and $y_2 \in
\ldiffmsigk$.\footnote{To understand what is going on here, simply note 
that if 
$(x_1 \in \lsigkmone \Leftrightarrow y_1 \not \in \ldelkmone)$ 
holds then by equation~$(**)$ we have $x \in
\overline{\ldiffmsigk} \Leftrightarrow y_2 \in \ldiffmsigk$.
Note also that
both of $x_1 \in \lsigkmone$ and $y_1 \not\in \ldelkmone$
can be very easily tested by a machine that has a $\sigmakmone$ oracle.}
This algorithm is not necessarily a $\diffmsigk$ algorithm, 
but it does inspire the following 
$\diffmsigk$ algorithm  to test whether $x \in \overline{\ldiffmsigk}$ in the
case that $x$ is an easy string for $p(|x|)$.

Let $L_1, L_2, \cdots, L_m$ be languages in $\Sigma^p_k$ such that
$\ldiffmsigk = L_1 - (L_2 - (L_3 - \cdots  (L_{m-1} - L_m) \cdots))$
and $L_1 \supseteq L_{2} \supseteq \cdots \supseteq L_{m-1}
\supseteq L_m$
(this can be done, as it is simply
the ``telescoping'' normal form of the levels of the 
boolean hierarchy over $\sigmak$, see
\protect\cite{cai-gun-har-hem-sew-wag-wec:j:bh1,hau:b:sets,wec:c:bh:ormaybe:wech:only:is:right}).
For $1 \leq r \leq m$, define $L_r'$ as the language accepted by the
following 
$\sigmak$ machine: On input $x$, guess $x_1$ with $|x_1| \leq p(|x|)$,
let $h(\pair{x_1, x}) = \pair{y_1, y_2}$,
and accept if and only if
$(x_1 \in \lsigkmone \Leftrightarrow y_1 \not \in \ldelkmone)$ and $y_2 \in
L_r$.

Note that $L'_r \in \sigmak$
for each $r$, and that $L'_1 \supseteq L'_{2} \supseteq \cdots
\supseteq L'_{m-1}  \supseteq L'_m$. 
We will show that if $x$ is an easy string for
length $p(|x|)$, then $x \in \overline{\ldiffmsigk}$ if and only if
$x \in L'_1 - (L'_2 - (L'_3 - \cdots  (L'_{m-1} - L'_m) \cdots))$.

So suppose that $x$ is an easy string for $p(|x|)$.
Define  $r'$
to be the unique
integer
such that
(a)~$0 \leq r' \leq m$,
(b)~$x \in L'_{s}$ for $1 \leq s \leq r'$,
and (c)~$x \not \in L'_{s}$ for $s > r'$.  It is immediate that  
$x \in L'_1 - (L'_2 - (L'_3 - \cdots  (L'_{m-1} - L'_m)
\cdots))$ if and only if $r'$ is odd.

Let $w$ be some string such that:
\begin{itemize}
\item $(\exists x_1 \in 
(\sigmastar)^{\leq p(|x|)}) (\exists y_1) [h(\langle x_1,
x \rangle ) = 
\langle y_1, w \rangle 
\wedge  (x_1 \in \lsigkmone \Leftrightarrow y_1 \not \in
\ldelkmone)]$, and
\item $w \in L_{r'}$ if $r' > 0$.
\end{itemize}
Note that such a $w$ exists, since $x$ is easy for $p(|x|)$. 
By the definition of $r'$ (namely, since $x \not\in L'_s$
for $s>r'$), 
$w \not \in L_s$ for all
$s > r'$. It follows that $w \in \ldiffmsigk$ if and only if $r'$ is odd.

It is clear, keeping in mind the definition of $h$, that $x 
\in \overline{\ldiffmsigk}$ iff $w \in \ldiffmsigk$, 
$w \in \ldiffmsigk$ iff 
$r'$ is odd, and
$r'$ is odd iff $x \in L'_1 - (L'_2 - (L'_3 - \cdots 
(L'_{m-1} - L'_m) \cdots))$. 
This completes the case where $x$ is easy,
as 
$L'_1 - (L'_2 - (L'_3 - \cdots 
(L'_{m-1} - L'_m) \cdots))$ in effect specifies 
a $\diffmsigk$ algorithm.

We say that $x$ is {\em hard for length $n$\/} if 
$|x| \leq n$ and $x$ is not easy for length $n$, i.e., if
$|x| \leq n$ and for all $x_1$ with $|x_1| \leq n$, $(x_1 \in \lsigkmone
\Leftrightarrow y_1 \in \ldelkmone)$, where
$h(\pair{x_1, x}) = \pair{y_1, y_2}$. Note that if $x$ is hard for
$p(|x|)$, then $x \not \in L'_1$.

If $x$ is a hard string for length $p(|x|)$, then $x$ induces a
many-one reduction from 
${\left(\lsigkmone\right)}^{\leq p(|x|)}$ to $\ldelkmone$,
namely, $f(x_1) = y_1$, where $h(\pair{x_1, x}) = \pair{y_1, y_2}$.
(Note that $f$ is computable in time 
polynomial in $\max(|x|,|x_1|)$.)
So it is not hard to see that if we 
choose $p$ appropriately large, then a hard string $x$ for
$p(|x|)$ induces $\Sigma_{k-1}^p$ algorithms for
$({L_1})^{=|x|}, ({L_2})^{=|x|}, \ldots, ({L_m})^{=|x|}$
(essentially since each is in $\sigmak  = \np^{\sigmakminusone}$,
$\lsigkmone$ is $\manyone$-complete for $\sigmakminusone$, and
$\np^{\deltakmone} = \sigmakminusone$),
which we can use to obtain a 
${\rm DIFF}_m(\Sigma_{k-1}^p)$  algorithm  for
${\ldiffmsigk}$, and thus certainly 
a $\diffmsigk$ algorithm for 
$\left(\overline{\ldiffmsigk}\right)^{=|x|}$.

However, there is a problem.  
The problem is that we cannot combine the $\diffmsigk$ algorithms for easy and
hard strings into one $\diffmsigk$ algorithm for
$\overline{\ldiffmsigk}$ that works all strings. Why?
It is 
too difficult to decide whether a string is easy or hard; to decide this
deterministically takes one query to $\sigmak$, and we cannot do that in
a $\diffmsigk$ algorithm. This is also the reason why the methods
from~\cite{hem-hem-hem:cOutByJour:downward-translation} 
failed to prove that if $\p \bolddelta \sigmatwo =  \np
\bolddelta \sigmatwo$, then $\sigmatwo  = \pitwo$. Recall from the
introduction that the latter theorem was proven  by Buhrman and
Fortnow~\cite{buh-for:t:two-queries}. We will use their technique at this
point. 
The following lemma, which we  will prove after we have finished the
proof of this theorem, states a generalized version of the technique
from~\cite{buh-for:t:two-queries}. It has been generalized to deal with
arbitrary levels of the polynomial hierarchy and 
to be useful in settings involving boolean
hierarchies.

\begin{lemma}\label{l-level}
Let $k>1$. 
For all $L \in \sigmak$, there exist a polynomial $q$ and a set
$\widehat{L} \in \pikmone$ such that
\begin{enumerate}
\item \label{conditionpart:foo} for each natural number $n'$, $q(n') \geq n'$, 
\item $\widehat{L} \subseteq \overline{L}$, and 
\item if $x$ is hard for $q(|x|)$, then $x \in \overline{L}$ iff $x \in
\widehat{L}$.
\end{enumerate}
\end{lemma}

We defer the proof 
of Lemma~\ref{l-level} until later in the paper, and we now 
continue with the proof of the current theorem.
From Lemma~\ref{l-level}, it follows that there exist sets $\widehat{L_1},
\widehat{L_2}, \ldots, \widehat{L_m} \in \Pi_{k-1}^p$ and polynomials $q_1,
q_2, \ldots, q_m$ with the following properties for all $1 \leq r \leq m$:
\begin{enumerate}
\item $\widehat{L_r} \subseteq \overline{L_r}$, and 
\item if $x$ is hard for $q_r(|x|)$, then $x \in \overline{L_r}$ iff $x \in
\widehat{L_r}$.
\end{enumerate}
Take $p$ to be an 
(easy-to-compute---we may without loss of 
generality require that there is an $\ell$
such that it is of the 
form $n^{\ell}+\ell$)
polynomial such that $p$ is at least as large as all the
$q_r$s, i.e., such that,
for each natural number $n'$, we have 
$p(n') \geq \max\{q_1(n'),\cdots,q_m(n')\}$. By 
the definition of hardness
and condition~\ref{conditionpart:foo} 
of Lemma~\ref{l-level}, 
if $x$ is hard for $p(|x|)$ then $x$ is hard
for $q_r(|x|)$ for all $1 \leq r \leq m$.  As 
promised earlier, we have now specified
$p$.  Define
${\lhatdiffmsigk}$ as follows: On input $x$, guess
$r$,
$r$ even,
$0 \leq r \leq m$, and accept if and only if
\begin{itemize}
\item $x \in L_r$ or $r = 0$, and
\item if $r < m$, then $x \in \widehat{L_{r+1}}$.
\end{itemize}
Clearly, ${\lhatdiffmsigk} \in \sigmak$. In addition, this set inherits certain
properties from the $\widehat{L_r}$s.  In particular, 
in light of the definition of $\lhatdiffmsigk$,
the definitions of the $\widehat{L_r}$s, and the fact
that
\begin{quote}
\noindent
$x \in \overline{\ldiffmsigk}$ iff for some even $r$, 
$0 \leq r \leq m$, we have:
$( x \in L_r$ or $r=0)$ and $( x \in \overline{L_{r+1}}$ or
$r=m)$,
\end{quote}
we have that the following properties hold:
\begin{enumerate}
\item ${\lhatdiffmsigk} \subseteq \overline{\ldiffmsigk}$, and 
\item if $x$ is hard for $p(|x|)$, then $x \in \overline{\ldiffmsigk}$ iff $x
\in
{\lhatdiffmsigk}$.
\end{enumerate}

Finally, we are ready to give the algorithm.  Recall that $L'_1, L'_2, \ldots
L'_m$ are sets in $\sigmak$ such that: 
\begin{enumerate}
\item  $L'_1 \supseteq L'_{2} \supseteq \cdots
\supseteq L'_{m-1}  \supseteq L'_m$, and 
\item if $x$ is easy for $p(|x|)$, then 
$x \in \overline{\ldiffmsigk}$ if and only if
$x \in L'_1 - (L'_2 - (L'_3 - \cdots  (L'_{m-1} - L'_m) \cdots))$, and
\item if $x$ is hard for $p(|x|)$, then $x \not \in L'_1$.
\end{enumerate}
We claim that for all $x$, $x \in \overline{\ldiffmsigk}$ iff 
$x \in (L'_1 \cup {\lhatdiffmsigk}) - (L'_2 - (L'_3 - \cdots
(L'_{m-1} - L'_m) \cdots))$, which completes the proof of
Theorem~\ref{t:keysubpart}, as $\sigmak$ is closed under union.

\begin{description}
\item[($\Rightarrow$)]  If $x$ is easy 
for $p(|x|)$, then $x \in 
 L'_1 - (L'_2 - (L'_3 - \cdots  (L'_{m-1} - L'_m) \cdots))$, and 
so certainly
$x \in (L'_1 \cup \lhatdiffmsigk) - (L'_2 - (L'_3 - \cdots
(L'_{m-1} - L'_m) \cdots))$.  If $x$ is hard 
for $p(|x|)$, then $x \in
{\lhatdiffmsigk}$ and $x \not \in L'_r$ for all $r$
(since $x \not\in L'_1$ and $L'_1 \supseteq L'_2 \supseteq \cdots$). 
Thus,
$x \in (L'_1 \cup {\lhatdiffmsigk}) - (L'_2 - (L'_3 - \cdots
(L'_{m-1} - L'_m) \cdots))$. 

\item[($\Leftarrow$)] Suppose $x \in (L'_1 \cup {\lhatdiffmsigk}) - (L'_2 -
(L'_3 -
\cdots (L'_{m-1} - L'_m) \cdots))$.
If $x \in {\lhatdiffmsigk}$, then $x \in
\overline{\ldiffmsigk}$.  If $x \not \in {\lhatdiffmsigk}$,
then $x \in  L'_1 - (L'_2 - (L'_3 - \cdots  (L'_{m-1} - L'_m) \cdots))$ and
so $x$ must be easy for $p(|x|)$ (as $x \in L'_1$, and this is 
possible only if $x$ is easy for $p(|x|)$).  
However, this says that $x \in \overline{\ldiffmsigk}$.~\qed
\end{description}

Having completed the proof of the theorem, we now 
return to the deferred proof of the lemma used within 
the theorem.

\smallskip

\noindent {\bf Proof of Lemma~\ref{l-level}.}
Let $L \in \sigmak$. We need to show that there exist a polynomial $q$ and a
set $\widehat{L} \in \pikmone$ such that
\begin{enumerate}
\item $\widehat{L} \subseteq \overline{L}$, and 
\item if $x$ is hard for $q(|x|)$, then $x \in \overline{L}$ iff $x \in
\widehat{L}$.
\end{enumerate}
From Definition~\ref{d:languages}, we know that:
$\lsigkprime$ is $\manyone$-complete for $\sigmak$, 
$\lsigkmone \in \Sigma_{k-1}^p$, 
$\lsigkmtwoprimeprime \in \Sigma_{k-2}^p$ , and
\begin{enumerate}
\item $\lsigkmone = \{\pair{{x},y,z} \condition |{x}| = |y|
\wedge (\exists z') [(|{x}| = |y| = |zz'|) \wedge
 \pair{{x}, y, zz'} \not \in \lsigkmtwoprimeprime]\}$, and
\item $\lsigkprime = \{{x} \condition
(\exists y \in \Sigma^{|{x}|}) (\forall z \in \Sigma^{|{x}|})
[\pair{{x},y,z} \in \lsigkmtwoprimeprime]\}$.
\end{enumerate}
Note that $\overline{\lsigkprime} = \{ {x} \condition
(\forall y \in \Sigma^{|{x}|}) (\exists z\in \Sigma^{|{x}|})
[\pair{{x},y,z} \not \in \lsigkmtwoprimeprime]\}$.

Since $L \in \sigmak$, and $\lsigkprime$ is $\manyone$-complete 
for $\sigmak$, there
exists a polynomial-time computable function $g$ such that,
for all $x$, $x \in
L$ iff $g(x) \in \lsigkprime$.

Let $q$ be such that 
(a)~$(\forall x \in \Sigma^n)(\forall y \in \Sigma^{|g(x)|})
(\forall z \in (\sigmastar)^{\leq |g(x)|})
[q(n) \geq |\pair{g(x),y,z}|]$ and 
(b)~$(\forall 
\widehat{m} \geq 0)[q(\widehat{m}+1) > 
q(\widehat{m}) > 0]$.  Note that we have ensured 
that for each natural number $n'$, $q(n') \geq n'$.

If $x$ is a hard string for length $q(|x|)$, then $x$ induces a
many-one reduction from 
${\left(\lsigkmone\right)}^{\leq q(|x|)}$ to $\ldelkmone$,
namely, $f_x(x_1) = y_1$, where $h(\pair{x_1, x}) = \pair{y_1, y_2}$.
(This is the $h$ from the proof of 
Theorem~\ref{t:keysubpart}.  One should treat the 
current proof as if it occurs immediately after the statement
of Lemma~\ref{l-level}.)
Note that $f_x$ is computable in time 
polynomial in $\max(|x|,|x_1|)$.

Let $\wh{L}$ be the language accepted by the following $\Pi_{k-1}^p$ 
machine:\footnote{\protect\label{foot:conondeterministic}{}For $k>1$, 
$\Pi_{k-1}^p = 
\conp^{\sigmakminustwo}$, and by a $\Pi_{k-1}^p$ machine 
we mean a co-nondeterministic machine with a $\sigmakminustwo$
oracle.  A co-nondeterministic machine by definition accepts
iff {\em all\/} of its computation paths are accepting 
paths.  (Some authors prefer requiring that all paths be 
rejecting paths;  the definitions are equivalent as long as one is 
consistent throughout regarding which model one is using.)}
\begin{tabbing}
On input $x$:\\
Compute $g(x)$\\
Guess $y$ such that $|y| = |g(x)|$\\
Set $w = \epsilon$ (i.e., the empty string)\\
While \= $|w| < |g(x)|$\\
\> if the $\deltakmone$ algorithm induced by $x$ for $\lsigkmone$
accepts
$\pair{g(x),y,w0}$\\
\> (that is, if $f_x(\pair{g(x),y,w0}) \in  \ldelkmone$),\\ 
\>then  $w = w0$ \\
\>else  $w = w1$\\
Accept if and only if $\pair{g(x),y,w} \not \in \lsigkmtwoprimeprime$.
\end{tabbing}

It remains to show that $\wh{L}$ thus defined fulfills the properties of
Lemma~\ref{l-level}. First note that the machine described above is clearly a
$\pikmone$ machine. To show that $\wh{L} \subseteq \overline{L}$, suppose that
$x \in \wh{L}$. Then (keeping in mind
the comments of footnote~\ref{foot:conondeterministic})
for every $y \in \Sigma^{|g(x)|}$, there exists a string
$w \in \Sigma^{|g(x)|}$ such that $\pair{g(x),y,w} \not \in
\lsigkmtwoprimeprime$.  This 
implies that $g(x) \in \overline{\lsigkprime}$, and thus 
that $x \in \overline{L}$.

Finally, suppose that $x$ is hard for $q(|x|)$ and that $x \in \overline{L}$.
We have to show that $x \in \wh{L}$. Since $x \in \overline{L}$,
$g(x) \in \overline{\lsigkprime}$. So,  $(\forall y \in
\Sigma^{|g(x)|}) (\exists z\in \Sigma^{|g(x)|}) [\pair{g(x),y,z} \not \in
\lsigkmtwoprimeprime]$.
Since $x$ is hard, $(\forall y \in \Sigma^{|g(x)|})\allowbreak
(\forall w \in (\sigmastar)^{\leq |g(x)|})
[\pair{g(x),y,w} \in \lsigkmone \Leftrightarrow 
f_x(\pair{g(x),y,w}) \in  \ldelkmone]$.
It follows that the algorithm above will find, for every $y \in
\Sigma^{|g(x)|}$,
a witness $w$ such that $\pair{g(x),y,w} \not \in
\lsigkmtwoprimeprime$, and thus the algorithm will accept $x$.~\qed

\section{A Downward Translation of Equality for Closure under
Complementation}

We now state our downward translation for closure under
complementation, Theorem~\ref{t:rest}.  
Theorem~\ref{t:rest} in part underpins our 
main result, Theorem~\ref{t:globalmain},
as Theorem~\ref{t:rest} is drawn on in the proof of 
Theorem~\ref{t:globalmain} (see the discussion immediately
after the statement of Theorem~\ref{t:globalmain}).
However, Theorem~\ref{t:globalmain} is not a corollary
of Theorem~\ref{t:rest};  the two results are 
incomparable.

\begin{theorem}\label{t:rest}
Let $m > 0$ and $0 < i < k - 1$.
If $\sigmai \bolddelta \diffmsigk$ is closed under complementation,
then $\diffmsigk = \codiffmsigk.$
\end{theorem}

{\bf Proof of Theorem~\ref{t:rest}} \quad 
Let $\lsigi$ and $\ldiffmsigk$ be $\manyone$-complete 
for $\sigmai$ and $\diffmsigk$ respectively.
Since $\lsigi \deltatilde \ldiffmsigk$ is $\manyone$-hard for
 $\sigmai \bolddelta \diffmsigk$ by Lemma~\ref{l:tildecomplete} 
(in fact, it is not hard to see that it even is $\manyone$-complete
for that class) and by
assumption
$\sigmai \bolddelta \diffmsigk$ is closed under complementation, there exists a
polynomial-time many-one reduction $h$ from
$\lsigi \deltatilde \ldiffmsigk$ to its complement.
That is, for all $x_1, x_2 \in \sigmastar$:
if $h(\pair{x_1, x_2}) = \pair{y_1, y_2}$, 
then:
$\pair{x_1, x_2} \in \lsigi \deltatilde \ldiffmsigk 
\Leftrightarrow \pair{y_1, y_2} \not \in
\lsigi \deltatilde \ldiffmsigk.$ Equivalently,
for all $x_1, x_2 \in \sigmastar$:
\begin{quotation}
\noindent
{\bf Fact~1:} \\
if $h(\pair{x_1, x_2}) = \pair{y_1, y_2}$, \\ 
then
$$(x_1 \in \lsigi \Leftrightarrow x_2 \notin \ldiffmsigk)
\mbox{ if and only if }
(y_1 \in \lsigi \Leftrightarrow y_2 \in \ldiffmsigk).$$
\end{quotation}

We can use $h$ to recognize some of $\overline{\ldiffmsigk}$ by a $\diffmsigk$
algorithm. 
In particular, 
we say that a string $x$ is {\em easy for length $n$\/} 
if there exists a string $x_1$ such that
$|x_1| \leq n$ and $(x_1 \in \lsigi
\Leftrightarrow y_1 \in \lsigi)$ where
$h(\pair{x_1, x}) = \pair{y_1, y_2}$.

Let $p$ be a fixed polynomial, which will be exactly 
specified later in the proof.  We have the following algorithm to test whether
$x \in \overline{\ldiffmsigk}$ in the 
case that (our input) $x$ is an easy string
for $p(|x|)$.
On input $x$, guess $x_1$ with $|x_1| \leq p(|x|)$, let
$h(\pair{x_1, x}) = \pair{y_1, y_2}$,
and accept if and only if
$(x_1 \in \lsigi \Leftrightarrow y_1 \in \lsigi)$  and
$y_2 \in \ldiffmsigk$.  
This algorithm is not necessarily a $\diffmsigk$ algorithm, 
but in the same way as in the proof of Theorem~\ref{t:keysubpart}, we can
construct sets $L'_1, L'_2, \ldots, L'_m \in \sigmak$ such that
if $x$ is an easy string for  length $p(|x|)$, then $x \in
\overline{\ldiffmsigk}$ if and only if
$x \in L'_1 - (L'_2 - (L'_3 - \cdots  (L'_{m-1} - L'_m) \cdots))$.

We say that $x$ is {\em hard for length $n$\/} if 
$|x| \leq n$ and $x$ is not easy for length $n$, i.e., if
$|x| \leq n$ and, for all $x_1$ with $|x_1| \leq n$, $(x_1 \in \lsigi
\Leftrightarrow y_1 \notin \lsigi)$, where
$h(\pair{x_1, x}) = \pair{y_1, y_2}$.

If $x$ is a hard string for length $n$, then $x$ induces a
many-one reduction from 
${\left(\lsigi\right)}^{\leq n}$ to $\overline{\lsigi}$,
namely, $f(x_1) = y_1$, where $h(\pair{x_1, x}) = \pair{y_1, y_2}$.
Note that $f$ is computable in time 
polynomial in $\max(n,|x_1|)$.

We can use hard strings to obtain a $\diff_m(\Sigma^p_{k-1})$ algorithm 
for $\ldiffmsigk$, and thus
(since $\diff_m(\Sigma^p_{k-1}) \subseteq \p^{\Sigma^p_{k-1}}
\subseteq \sigmak \cap \pik$) 
certainly a $\diffmsigk$ algorithm for
$\overline{\ldiffmsigk}$.
Let $L_1, L_2, \cdots, L_m$ be languages in $\Sigma^p_k$ such that
$\ldiffmsigk = L_1 - (L_2 - (L_3 - \cdots  (L_{m-1} - L_m) \cdots))$.
For all $1 \leq r \leq m$, let $M_r$ be a
$\Sigma_{k-i}^p$ machine 
such that $M_r$ with oracle $\lsigi$
recognizes  $L_r$.
Let the run-time of all $M_r$s be bounded by polynomial $p$,
which without loss of generality is easily computable 
and satisfies
$(\forall 
\widehat{m} \geq 0)[p(\widehat{m}+1) > 
p(\widehat{m}) > 0]$ (as promised earlier, we have
now specified $p$).
Then, for all $1 \leq r \leq m$,
$${\left( L_r \right)}^{= n} = 
\left( 
L 
\left(M_r^{   {\left( \lsigi
\right)}^{\leq p(n)}
} 
\right) 
\right)^{= n}.$$
If there exists a hard string for length $p(n)$, then that hard string
induces a reduction from $\left(\overline{\lsigi}\right)^{\leq p(n)}$ to
$\lsigi$.

Let $L \in \Sigma^p_{i-1}$ and $r$ be a polynomial such that
$r$ is easily computable and, for all $x$, 
$x\in \lsigi$ iff 
$(\exists y \in (\sigmastar)^{\leq r(|x|)})[\pair{x,y}\not\in L]$.

We will show that with
any hard string for length $p(n)$ in hand, call it $w_n$, 
there exist $\Sigma_{k-1}^p$ algorithms for $(L_1)^{=n}, 
(L_2)^{=n}, \ldots, (L_m)^{=n}$.

Let $\widehat{M_r}$ be the following $\Sigma^p_{k-i}$ machine.
On input $x$ of length $n$, $\widehat{M_r} (x)$ simulates the work 
of $M_r(x)$ until $M_r(x)$ asks a query, 
call it $q$. Then $\widehat{M_r}$ guesses whether this query will be answered
``Yes'' or ``No.'' If $\widehat{M_r}$ guesses ``Yes,'' then $\widehat{M_r}$
guesses a certificate
$y \in (\sigmastar)^{\leq r(|q|)}$, makes
the query $\pair{q,y}$ to $L$, rejects if the
answer is ``Yes,'' and proceeds with 
the simulation if the answer is ``No.''
If $\widehat{M_r}$ guesses that the answer to $q$ is ``No,'' then
$\widehat{M_r}$ guesses that $q \not \in \lsigi$, or in other words that $q \in
\overline{\lsigi}$. Now we will use the reduction from $\overline{\lsigi}$ to
$\lsigi$, because
$q \in \overline{\lsigi}$ if and only~if 
the first component of $h(\pair{q,w_n})$ is in $\lsigi$.
Let $q'$ denote the first component of $h(\pair{q,w_n})$.
$\widehat{M_r}$ also guesses  $y' \in 
(\sigmastar)^{\leq r(|q'|)}$, makes the query 
$\pair{q',y'}$ to $L$,  rejects if the
answer is ``Yes,'' and proceeds with 
the simulation if the answer is ``No.''
Clearly, $\widehat{M_r}$ is a $\Sigma^p_{k-i}$ machine that recognizes 
$(L_r)^{=n}$ with queries to a $\Sigma^p_{i-1}$ oracle, namely $L$.

It follows that if there exists a hard string for length $p(n)$, then
this string induces a $\diff_m({\Sigma_{k-1}^p})$ algorithm for 
${\left({\ldiffmsigk}\right)}^{=n}$, and thus certainly
a $\diffmsigk$ algorithm for
${\left(\overline{\ldiffmsigk}\right)}^{=n}$.

It follows that there exist $m$ $\sigmak$ sets, say, $\widehat{L_r}$ for
$1 \leq r \leq m$, such that the following holds:
For all $x$, if $x$ (functioning as $w_{|x|}$ above)
is a hard string for length $p(|x|)$, then
$x \in \overline{\ldiffmsigk}$ if and only if
$x \in \widehat{L_1} - (\widehat{L_2} - (\widehat{L_3} - \cdots 
(\widehat{L_{m-1}} - \widehat{L_m}) \cdots))$.

However, now we have an outright $\diffmsigk$ algorithm 
for $\overline{\ldiffmsigk}$: For $1 \leq r \leq m$ define a
$\np^{\Sigma_{k-1}^p}$  machine  $N_r$ as follows:
On input $x$, the NP base machine of $N_r$ 
executes the following algorithm:
\begin{enumerate}
\item Using its 
$\Sigma_{k-1}^p$ oracle, it deterministically 
determines whether the input $x$ is an easy string for 
length $p(|x|)$.  This can be done, as checking whether 
the input is an easy string for length $p(|x|)$ can be done 
by two queries to $\Sigma_{i+1}^p$, and $i+1 \leq k-1$ by 
our $i < k-1$ hypothesis.
\item If the previous step determined that the input is not 
an easy string, then the input must be a hard string
for length $p(|x|)$.  
So simulate the $\sigmak$ algorithm for $\widehat{L_r}$
induced by this hard string
(i.e., the input $x$ itself) on input $x$ (via our NP
machine itself simulating the base level of the 
$\sigmak$ algorithm and using the NP machine's oracle to 
simulate the oracle queries made by the base level NP machine 
of the 
$\sigmak$ algorithm being simulated).
\item If the first step determined 
that the input $x$ is easy for length $p(|x|)$, then our NP
machine
simulates (using itself and its oracle) 
the $\sigmak$ algorithm for  $L'_r$ on input $x$.
\end{enumerate}
It follows that, for all $x$, $x  
\in \overline{\ldiffmsigk}$ if and only if
$x \in L(N_1) - (L(N_2) - (L(N_3) - \cdots 
(L(N_{m-1}) - L(N_m)) \cdots))$.
Since $\overline{\ldiffmsigk}$ is complete for $\codiffmsigk$,
it follows that  $\diffmsigk = \codiffmsigk$.~\qed

An underlying goal of this paper is to show that 
Theorem~\ref{t:hhh-nonBHdownward} holds even for $k=2$ 
and the bounded query hierarchy (that is, that 
Corollary~\ref{c:special-case} holds),
and Theorem~\ref{t:rest} plays a central role
in establishing this.  We mention that---though it is in no way 
needed to establish Corollary~\ref{c:special-case}, and its proof is 
somewhat less transparent and more
technical than that of Theorem~\ref{t:rest}---it is possible
to prove a slightly stronger version of 
Theorem~\ref{t:rest}
that removes the asymmetry in its statement:
Let $s,m > 0$ and $0 < i < k - 1$.
If $\diffssigi \bolddelta \diffmsigk$ is closed under complementation,
then $\diffmsigk = 
\codiffmsigk$~\cite{hem-hem-hem:inprep:diff-diff-new-downward}.

\section{Conclusions}\label{s:conclusions}

We have proven the following downward translations of equality.
\begin{enumerate}
\item Let $m > 0$ and 
$0 < i < k$. If $\deltai \bolddelta \diffmsigk = \sigmai
\bolddelta \diffmsigk$, then
$\diffmsigk = \codiffmsigk$.
\item
Let $m > 0$ and $0 < i < k - 1$.
If $\sigmai \bolddelta \diffmsigk$ is closed under complementation,
then $\diffmsigk = \codiffmsigk$.
\end{enumerate}
As mentioned in the introduction, these results 
extend
the polynomial 
hierarchy's previously known downward translations
of equality.  More importantly, 
they
show that Theorem~\ref{t:hhh-nonBHdownward}
can be extended to all $k>1$ cases 
even for each level of the bounded
query hierarchy.
\begin{corollary}\label{c:special-case}
For each $m > 0 $ and each $k>1$ it holds that:
$$ \psigkmqueries = \psigkmplusone \implies \diffmsigk = \codiffmsigk.$$
\end{corollary}

\typeout{ADD THIS IN JOURNAL/TR VERSION BUT FOR SPACE REASONS LEAVE OUT}
\typeout{OF CONFERENCE.  ``THIS'' MEANS SOME COMMENTED OUT TEXT... SEE}
\typeout{THE SOURCE LATEX CODE AT THIS LOCATION!!!}

Corollary~\ref{c:special-case} itself has an interesting further
consequence.  From this corollary, it follows 
(for exactly the reasons discussed 
in~\cite{hem-hem-hem:jtoappear:downward-translation}) 
that for a number of
previously missing cases (namely, when $m>1$ and $k=2$), the
hypothesis $ \psigkmqueries = \psigkmplusone$ implies that the
polynomial hierarchy collapses to about one level lower in the
boolean hierarchy over $\sigmakplusone$ than could be concluded
from previous papers.  In particular, one can now conclude that,
for all cases where $m>0$ and $k>1$,
$ \psigkmqueries = \psigkmplusone$ implies that 
each set
in the polynomial hierarchy can be accepted by a~P machine that makes
$m-1$ truth-table queries to $\Sigma^p_{k+1}$, and that in addition
makes one query to $\Delta^p_{k+1}$ (in fact, a bit more 
can be claimed, see~\cite{hem-hem-hem:inprep:easy-hard-survey}).

{\singlespacing

\small

\bibliography{gry}

}

\clearpage

\end{document}